\documentclass[english]{emulateapj}
\usepackage[T1]{fontenc}
\usepackage[latin1]{inputenc}
\usepackage{graphicx}
\usepackage{amssymb}

\makeatletter
\usepackage{subfigure}
 \usepackage{multirow} 
\usepackage{times}
\usepackage{amsmath}

\newcommand{\yr}{{\,\rm yr}}

\newcommand{\Myr}{{\,\rm Myr}}

\newcommand{\pc}{\,\mathrm{pc}}

\newcommand{\Mimbh}{M_{\rm imbh}}
\newcommand{\eimbh}{e_{\rm imbh}}
\newcommand{\aimbh}{a_{\rm imbh}}

\newcommand{\Mbh}{M_{\bullet}}

\newcommand{\Mo}{M_{\odot}}

\newcommand{\rad}{\rm rad}

\shorttitle{SECULAR DYNAMICAL ANTI-FRICTION}
\shortauthors{MADIGAN AND LEVIN}

\usepackage{babel}
\makeatother
\begin{document}
\bibliographystyle{apj} 

\title{Secular Dynamical Anti-Friction in Galactic Nuclei}

\author{Ann-Marie Madigan$^1$ and Yuri Levin$^{1,2}$}
\affiliation{$^1$Leiden Observatory, Leiden University, P.O. Box 9513, NL-2300 RA Leiden, The Netherlands \\$^2$School of Physics, Monash University, Clayton, Victoria 3800, Australia}
\email{madigan@strw.leidenuniv.nl}

\begin{abstract}

We identify a gravitational-dynamical process in near-Keplerian potentials of galactic nuclei that occurs when an intermediate-mass black hole (IMBH) is migrating on an eccentric orbit through the stellar cluster towards the central supermassive black hole. We find that, apart from conventional dynamical friction, the IMBH experiences an often much stronger systematic torque due to the secular (i.e., orbit-averaged) interactions with the cluster's stars. The force which results in this torque is applied, counterintuitively, in the same direction as the IMBH's precession and we refer to its action as ``secular dynamical anti-friction'' (SDAF).  We argue that SDAF, and not the gravitational ejection of stars, is responsible for the IMBH's eccentricity increase seen in the initial stages of previous $N$-body simulations. Our numerical experiments, supported by qualitative arguments, demonstrate that (1) when the IMBH's precession direction is artificially reversed, the torque changes sign as well, which decreases the orbital eccentricity; (2) the rate of eccentricity growth is sensitive to the IMBH migration rate, with zero systematic eccentricity growth  for  an IMBH whose orbit is artificially prevented from inward migration; and (3) SDAF is the strongest when the central star cluster
is rapidly rotating. This leads to eccentricity growth/decrease for the clusters rotating in the
opposite/same direction relative to the IMBH's orbital motion. \\

\end{abstract}




\section{Introduction}

Recent studies on secular dynamics in near-Keplerian potentials \citep{Rau96,Hop06a,Eil09,Koc11,Mad11b} have focused attention on the long-term dynamical evolution of stars and compact objects in galactic nuclei. Due to persistent gravitational torques exerted by stellar orbits on each other as they precess slowly around the supermassive black hole (SMBH), the angular momentum evolution of stellar orbits can proceed at a much higher rate than that of energy evolution \citep[resonant relaxation; ][]{Rau96}. In this paper we extend this research to study secular dynamical effects on the inspiral of a massive body, e.g., an intermediate-mass black hole (IMBH), into the combined potential of an SMBH and nuclear stellar cluster.  

As a body of mass $M$ moves through a distribution of field stars of individual mass $m_a$, it experiences a frictional force anti-parallel to its velocity ${\bf v}$,
\begin{equation}
M \frac{d{\bf v}}{dt} = - 16 \pi^2 G^2 M^2 m_a \ln \Lambda \left[ \int_0^v dv_a v_a^2 f(v_a) \right] \frac{{\bf v}}{v^3}.
\label{eq:chandra}
\end{equation}
This is Chandrasekhar's dynamical friction formula \citep{Cha43,Tre84} for an isotropic field star distribution normalized such that the density $n(r) = \int d^3v_a f(r,v_a)$, and $\ln \Lambda$ is the Coulomb logarithm. The formula accounts only for stars moving slower than the massive body, and it neglects self-gravity of the stars themselves; \citet{Ant12} have recently updated the formula to include changes induced by the massive body on the stellar velocity distribution and the contribution from stars moving faster. The stars are deflected by the massive body into a gravitational ``wake'' behind it which results in an increase in stellar density, the amplitude of which is proportional to $M$ \citep{Dan57,Kal71,Mul83}; hence, the gravitational force experienced by the massive body is proportional to $M^2$. This frictional force acts to decrease the kinetic energy and angular momentum of the massive body and it sinks towards the center of the potential. 

The picture described above does not take into account the orbit-averaged dynamics that is particular for near-Keplerian potentials and that has been shown to play a central role in the angular momentum evolution of stars near the SMBH. Indeed, it is natural to assume that some other form of dynamical friction could be associated with the orbit-averaged
potential created by the IMBH, as this potential is rotating around the SMBH due to precession of the IMBH's eccentric orbit. The possibility of this secular dynamical friction was already suggested by \citeauthor{Rau96} in 1996. In this paper we explore this effect and find that, contrary to our experience with ordinary dynamical friction, the resulting torque comes from a gravitational force acting in the same direction as that of the IMBH precession.\footnote{This is true if the stellar cluster is non-rotating. As we will see below, if the cluster is rapidly rotating, then the torquing gravitational force can be applied in the same direction as the cluster's rotation.} We therefore refer to this effect as ``secular dynamical anti-friction'' (SDAF). 

For the remainder of the paper we will keep referring to the massive in-spiraling body as an IMBH, though the reader should keep in mind that the essential dynamics will hold for any massive body moving in a near-Keplerian potential. As large mass ratio massive black hole binaries are expected to coalesce in at least some merging and non-merging galaxies \citep{MilMer03,Pre11}, the dynamics of such systems are important to understand. 

\pagebreak

In particular, IMBH inspirals into SMBHs will be a major source of gravitational waves for space-based laser interferometers such as the proposed European New Gravitational Wave Observatory \citep{Amo12}. Simulations of the inspiral of IMBHs ($10^2 \! \-- \!10^4 \Mo$) through nuclear stellar clusters embedding an SMBH have shown an increase in eccentricity of the IMBH \citep{Bau06,Mat07,Loc08,Iwa11,Ses11,Ant12,Mei11}. Here, we show that the theory of SDAF can explain the initial stages of this phenomenon and present results of $N$-body simulations set up to test this claim. 

\section{Secular Dynamical Anti-Friction}

\begin{figure}[t!]
	\begin{center}
			\includegraphics[height=85 mm, angle=270]{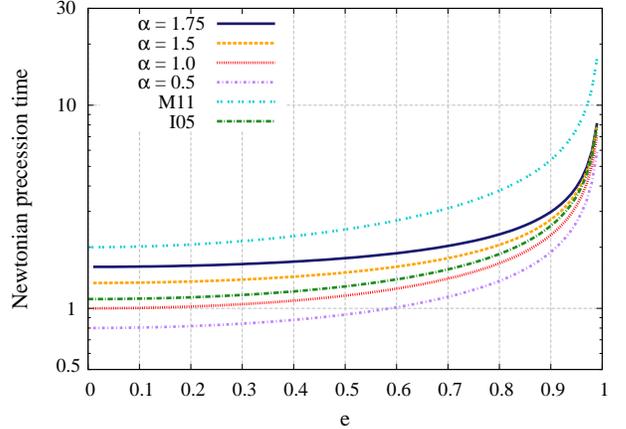}
		\caption{Dependence of the precession time ($\Mbh /( N_< m) P(a) = 1$) of a stellar orbit on orbital eccentricity, $e$. Several values are plotted for different power-law density profiles, $n(r) \propto r^{-\alpha}$. The dependence given in \citet[][M11]{Mer11a} and \citet[][I05]{Iva05} are plotted for comparison.
		 \label{f:prec_e}}
	\end{center}
\end{figure}

Let us consider the combined gravitational potential of an SMBH, mass $\Mbh$, embedded in a nuclear star cluster with a power-law number density distribution, $n(r) \propto r^{-\alpha}$, and individual stellar masses $m$. The specific energy and angular momentum of a stellar orbit can be written in terms of Kepler elements as
\begin{equation}
E = - \frac{G\Mbh}{2 a}, 
\end{equation}
\begin{equation}
|{\bf J}| = |{\bf r} \times {\bf v}| = [{G\Mbh a}({1 - e^2})]^{1/2},
\end{equation}
where $a$ is the semi-major axis of the elliptical orbit, and $e$ is the eccentricity. The eccentricity vector, ${\bf e}$, of the star is that which points from the occupied focus of the orbit to the periapsis of the orbit and has a magnitude equal to the scalar eccentricity of the orbit,
\begin{equation}
{\bf e} = \frac{1}{G \Mbh} ({\bf v} \times {\bf J}) - \hat{\bf r}.
\end{equation}
The orbit of a star in this potential is nearly Keplerian (i.e., closed) but ${\bf e}$ will precess due to the additional (Newtonian) potential and general relativity. We define the precession time, $t_{\rm prec}$, as a timescale over which an orbit precesses by $2\pi ~ \rad$ in its plane. As Newtonian precession acts with retrograde motion (i.e., in the direction opposite of a star's motion) and general relativistic precession with prograde motion,
\begin{equation}\label{eq:tprec}
t_{\rm prec} = \left|\frac{1}{t^{\rm cl}_{\rm prec}} - \frac{1}{t^{\rm GR}_{\rm prec} }\right|^{-1} .
\end{equation}
The general relativistic precession time is given by
\begin{equation}
t^{\rm GR}_{\rm prec}  = \frac{a(1 - e^2) c^2}{3 G \Mbh} P(a),
\end{equation}
with the orbital period $P(a) = 2 \pi \sqrt{a^3/G\Mbh}$. The Newtonian precession time due to the addition of the star cluster mass in an otherwise-Keplerian potential, $t^{\rm cl}_{\rm prec}$, can be written for $\alpha \ne 2$ as
\begin{equation}\label{eq:tM}
t^{\rm cl}_{\rm prec} = \pi (2 \!-\! \alpha) f(e, \alpha)^{-1} \left[ {\Mbh\over N_< m}P(a) \right],
\end{equation}
where $N_<$ is the number of stars within semi-major axis $a$, and 
\begin{equation}
f(e, \alpha) = \frac{\partial}{\partial \mathcal{J}} \left( \frac{1}{\mathcal{J}} \int^{\pi}_0 \left[ \frac{\mathcal{J}^2}{1 + \sqrt{1 - \mathcal{J}^2} \cos{\phi}} \right]^{4 - \alpha} d\phi \right)
\end{equation}
\citep{Lan69}. Here $\mathcal{J} = J/(G\Mbh a)^{1/2} = ({1 - e^2})^{1/2}$. For all values of $\alpha$, the Newtonian precession time of a stellar orbit increases with eccentricity. The difference in precession times for circular and highly eccentric stellar orbits varies with $\alpha$. We plot $t_{\rm prec}$ in Figure \ref{f:prec_e}, setting the quantity in brackets $[\Mbh /( N_< m) P(a)]$ in Equation (\ref{eq:tM}) equal to 1, with $f(e, \alpha)$ as given in \citet{Mad11b}. For a steep cusp, $\alpha = 1.75$ \citep{Bah76}, there is a factor of $\sim \! 5$ difference in precession time for a near-circular ($e = 0.01$) and near-radial ($e = 0.99$) orbit, which increases to $\sim \!7$ for a shallow profile $\alpha = 0.5$. We plot Equation (13) from \citet{Mer11b} and Equation (A11) from \citet{Iva05} ($\alpha = 1.5$) as a comparison. Differences arise in the normalization of the functions due to approximations made in deriving the analytical formulae in these papers; the monotonic dependence of the precession time on orbital eccentricity persists. We note that the exact expressions for the precession time in \citet{Iva05} agrees with numerical evaluation of our Equation (\ref{eq:tM}).

We now introduce an IMBH of mass $\Mimbh$ on an orbit with non-zero eccentricity, $\eimbh$, and semi-major axis, $\aimbh$, within the dynamical radius, $r_h$, of the SMBH, defined such that the mass in stars at $r_h$ equals that of the SMBH, $N(<r_h) m = \Mbh$. As we are interested in secular dynamics,\footnote{The term ``secular'' refers to long-period dynamics, in which the dependence on the mean longitude of an orbit is dropped from the disturbing function \citep[see, e.g.,][]{Der99}.} it is useful to envisage the mass of the IMBH spread smoothly out along its orbit, such that the local density at any segment is inversely proportional to its local velocity. The position of furthest distance from the SMBH, the apoapsis of the orbit, will therefore contain the most mass. Over a timescale longer than its orbital period but much less than the precession time, it will exert a strong (specific) gravitational torque on the orbit of a star with a similar semi-major axis \citep{Gur07} on the order of
\begin{equation}
|{\bf \tau}| = |{\bf r} \times {\bf F}| \sim \frac{G \Mimbh \eimbh e}{\aimbh} \delta \phi,
\end{equation}
where $\delta \phi$ is the angle between two orbits. 
\begin{figure*}[t!]
  \centering
  \subfigure{
    \includegraphics[trim=2cm 0cm 2cm 0cm, clip=true, angle = 0, scale=0.23]{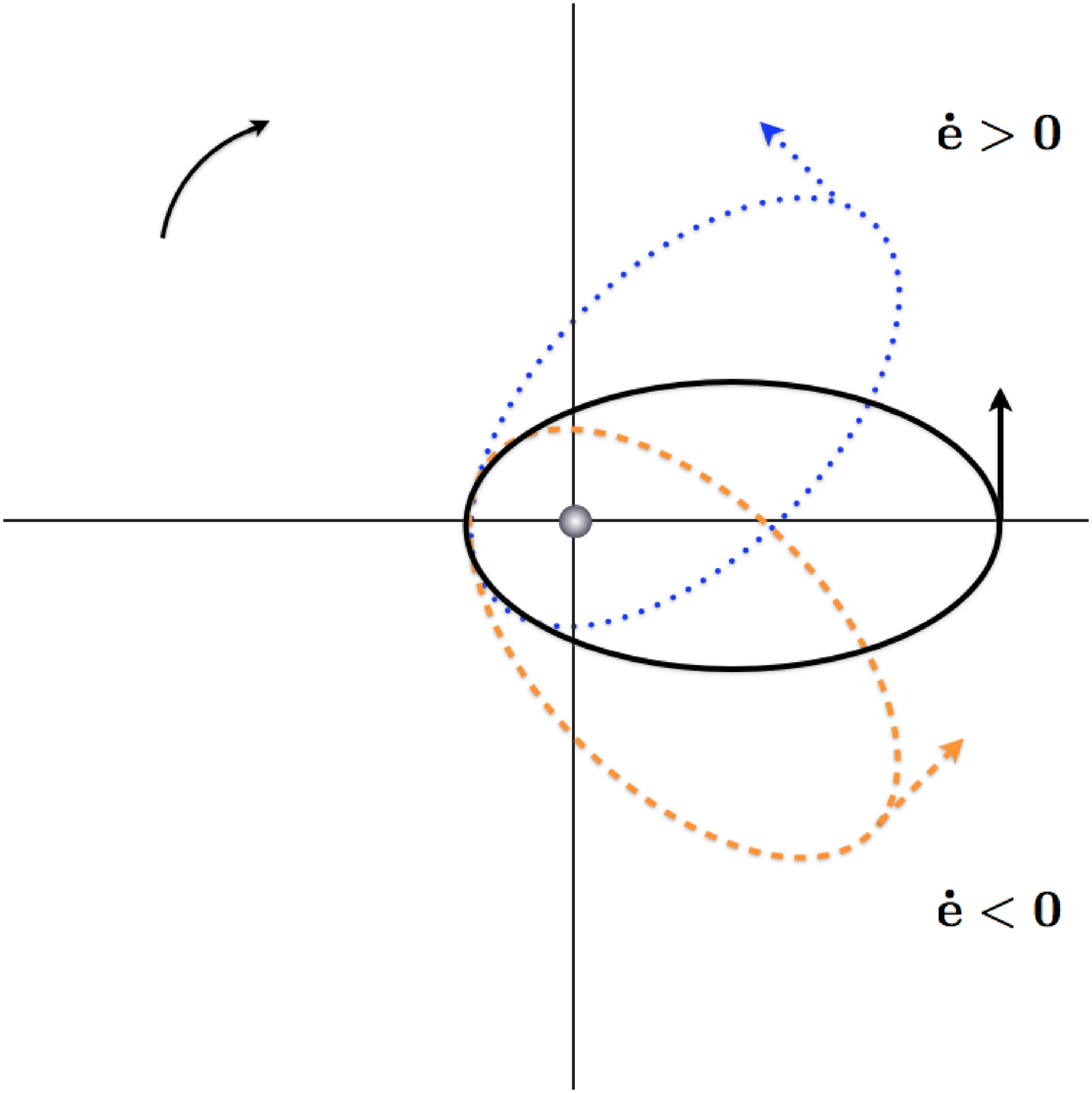}
  }
  \subfigure{
    \includegraphics[trim=2cm 0cm 2cm 0cm, clip=true, angle = 0, scale=0.23]{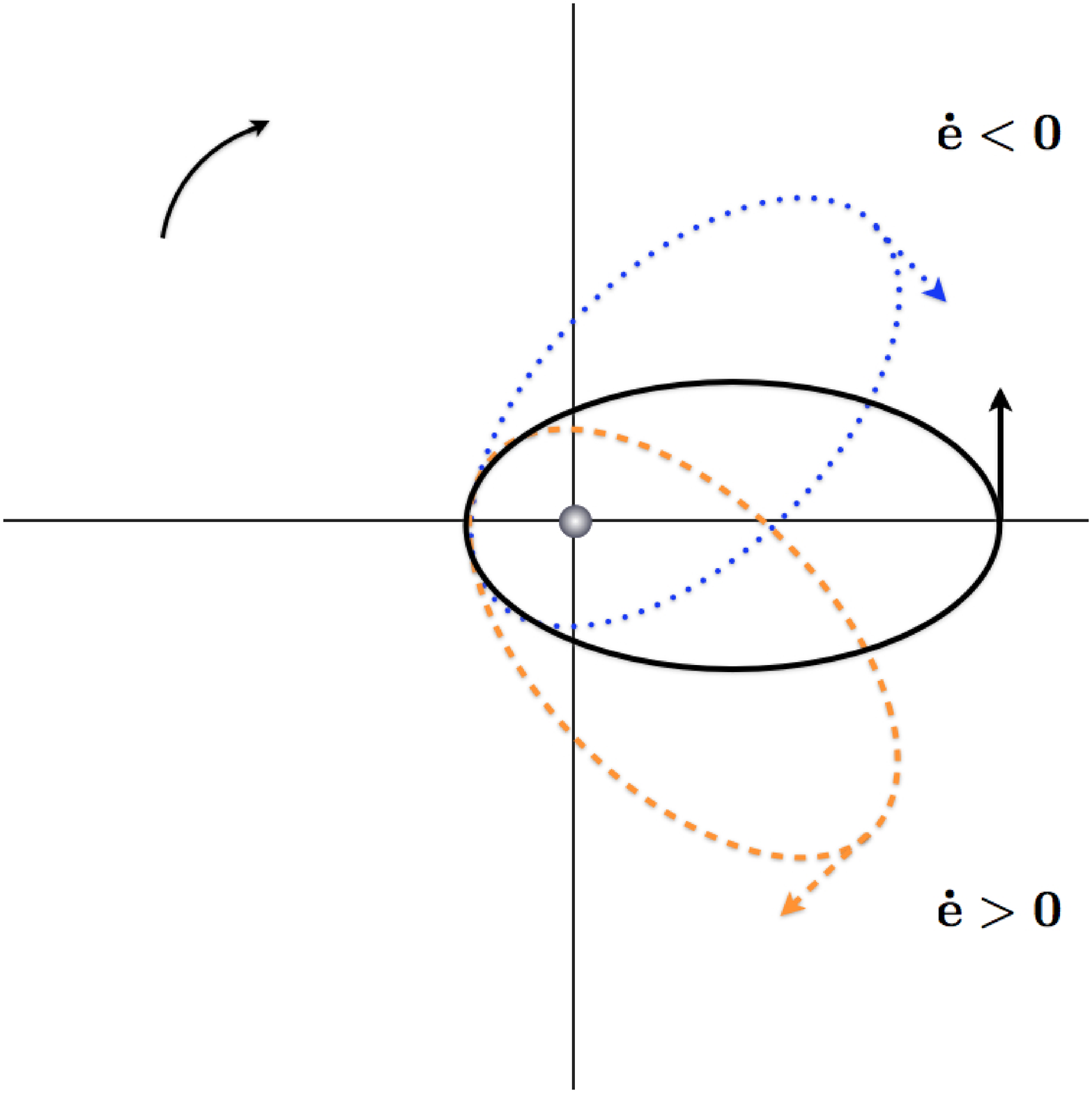}
  }
		\caption{Left: schematic of orbits of IMBH (black, filled line) and prograde leading (orange, dashed line) and prograde trailing (blue, dotted line) stellar orbits in the frame of reference rotating with the IMBH orbit. The velocity vectors of each are indicated at apoapsis and the direction of precession of the IMBH is shown with a curved arrow at top left. Right: schematic of orbits of IMBH (black, filled line) and retrograde leading (orange, dashed line) and retrograde trailing stellar orbits (blue, dotted line) in the frame of reference rotating with the IMBH orbit.		
		 \label{f:scheme}}
\end{figure*}

\pagebreak 

The orbit of the IMBH will not affect the energy (or equivalently, the semi-major axis) of the stellar orbit as its gravitational potential is stationary over this timescale. A star on a coplanar orbit with the IMBH will experience a torque that is parallel (or anti-parallel) to its angular momentum vector ${\bf J}$. 

Hence, the torque will act to change the magnitude of ${\bf J}$; equivalently, the eccentricity of the orbit. Let us consider the stellar orbit with an eccentricity $e \sim \eimbh$ and semi-major axis $a \sim \aimbh$. We place the orbit such that its eccentricity vector forms an angle $ \delta \phi$ with that of the IMBH. 

If  there existed no gravitational interaction (secular or otherwise) between the IMBH and the star, they would precess in the background stellar potential at the same rate. Their eccentricity vectors would rotate by $2 \pi ~\rad$ in one $t_{\rm prec}$ and $\delta \phi$ would remain constant. Switching on secular gravitational interactions results in a strong secular encounter between the star and the IMBH. The star's orbit is pushed away from the IMBH, transferring negative or positive angular momentum to the orbit of the IMBH, depending on its orientation. We elaborate on the dynamics in the following.

If the star is moving on a prograde orbit (in the direction of motion of the IMBH) that leads in front of the IMBH (in terms of precession direction; see Figure \ref{f:scheme}: left), it will experience a torque in the direction of its angular momentum vector ${\bf J}$ such that $dJ/dt > 0$. Due to the resulting decrease in eccentricity, the stellar orbit will precess faster (see Figure \ref{f:prec_e}) and move away from the IMBH.\footnote{This mechanism involves the same interplay between stellar torques and orbital precession which results in an instability in an eccentric disk of stars near an SMBH \citep{Mad09}.} Conversely if the stellar orbit is trailing, the torque exerted on it will be anti-parallel to ${\bf J}$. The stellar orbit will increase in eccentricity, precess more slowly and in effect be pushed away from the orbit of the IMBH. In both cases,  $d|\delta\phi|/dt > 0$.

A star moving along a retrograde orbit (see Figure \ref{f:scheme}: right) precesses in the opposite direction to that of the IMBH. If the orbit is leading in front of the IMBH, the torque will act to increase its eccentricity and hence decrease the differential precession rate between the two orbits. This increases the timescale over which the stellar orbit experiences a coherent torque and its eccentricity can increase to one. In this scenario the orbit can easily ``flip'' through $e = 1$ to a prograde orbit, and be pushed away from the IMBH as described above. If if does not flip, the retrograde orbit will precess past the IMBH orbit, experience a torque which will decrease its eccentricity and hence will increase the differential precession rate between the two orbits. Once again, in both cases the stellar orbit is pushed away from that of the IMBH, $d|\delta\phi|/dt > 0$. 

In contrast to Chandrasekhar's dynamical friction, which acts to increase the density of stars behind a massive body, the effect of SDAF is to push stellar orbits away from that of the massive body. The varying strength of Chandrasekhar's dynamical friction on different segments of a massive body's orbit causes evolution of orbital eccentricity, the sign of which is dependent on the background stellar density profile \citep[e.g.,][]{Gou03,Lev05}. With SDAF, it is the relative number of prograde/retrograde stellar orbits which determines the sign of the orbital eccentricity evolution of the IMBH. 

\subsection{Increase in orbital eccentricity}

During a strong secular encounter of the IMBH with a star, the angular momentum of the star is randomized. The rate of encounters of the IMBH with retrograde stars is higher than that of prograde stars, and the net result is to decrease the angular momentum of the IMBH. If the precession direction of the IMBH is artificially reversed, the net torque changes sign and the angular momentum of the IMBH increases.
Migration is important in this picture. If the IMBH does not migrate, then the relative number of prograde and retrograde orbits quickly adjusts so there is no net secular torque in the steady state. However, if the IMBH migrates due to dynamical friction, the adjustment does not have time to occur and there is an anti-friction torque. In the next section, we perform numerical simulations to test this picture. 

\subsection{Comparison with theories in the literature}

The $N$-body simulations of \citet{Iwa11} show an increase in the eccentricity of an IMBH as it spirals into an SMBH. With careful analysis of the orbital properties of the field stars, the authors attribute the increase to secular chaos (and the Kozai mechanism), brought about by the non-axisymmetric potential induced by the IMBH. 

\pagebreak 
This acts to bring the relative number of orbits back into balance after prograde-moving stars are preferentially ejected by the IMBH (retrograde stars have larger relative velocities with respect to the IMBH than prograde, and hence are ejected less frequently). The retrograde orbits moving to prograde orbits extract angular momentum from the orbit of the IMBH and its eccentricity increases.

\citet{Ses11} present both hybrid\footnote{The hybrid model couples numerical three-body scatterings to an analytical formalism for interactions between the cusp and the IMBH \citep{Ses08b}.} and direct $N$-body simulations with stellar cusps of varying degrees of rotation, to examine the importance of the relative number of prograde/retrograde orbits on the orbital evolution of the IMBH. The authors propose a hypothesis in which the ejections of stars, via the slingshot mechanism, are the cause of the evolution of the IMBH eccentricity. Most retrograde orbits which end up being ejected from the potential, donate their negative angular momentum to the IMBH and move to prograde orbits before they escape. Again it is the preferential ejection of prograde orbits that drives the increase of the orbital eccentricity of the IMBH.

With SDAF, it is the cumulative secular torques from retrograde stellar orbits that lead to the increase in orbital eccentricity of an IMBH, irrespective of ejections of stars from the potential. SDAF makes testable predictions which can distinguish itself from the above theories.
\begin{enumerate}
\item If the orbit of the IMBH is made to artificially precess in the opposite direction to its true motion, its rate of change of angular momentum should reverse sign as well.

\item If escaping stars are injected back into the cluster with the same parameters as before their interaction with the IMBH, the eccentricity evolution of the IMBH should not qualitatively be affected.

\item Due to the ``pushing away'' of coplanar stellar orbits, there will be a density anisotropy in the stellar distribution as a result of SDAF. 

\item If dynamical friction is artificially slowed, there will be a lower rate of axisymmetric flow of retrograde orbits to one side of the IMBH's orbit and hence a lower negative rate of change of angular momentum.
\end{enumerate}

 In the following section we present the results of numerical trials to test the theory of SDAF and decrease of orbital angular momentum of an IMBH in a galactic nucleus.

\section{Numerical Method and Simulations}

The simulations presented in this paper are performed using a special-purpose, mixed-variable symplectic $N$-body integrator as described in \citet{Mad11b}. We use a galactic nucleus model with parameters chosen to represent the Galactic center. At the center of the coordinate system, there is an SMBH with $\Mbh = 4 \times 10^6 \Mo$ \citep{Ghe03a}, and an IMBH with a mass $\Mimbh$ within the range $[10^3 \!-\!10^4] \Mo$ with $0.04 \pc < a < 0.08 \pc$. We simulate the nuclear star cluster as a smooth potential with a power-law number density profile $n(r) \propto r^{-\alpha}$; this decreases the computation time and lends a clear, direct interpretation to our results. Typically $\alpha \in [0.5,1.75]$, and is normalized such that at one parsec the stellar mass is $M(< 1\pc)= 1 \times 10^6 \Mo$ \citep{Sch07, Tri08, Sch09}. To test our theory of SDAF, we include an isotropically distributed population of test stars with individual masses $m$ within the range $[10\!-\!100] \Mo$ which gravitationally interact only with the IMBH, not with each other. In doing so we isolate the effects of the IMBH on the stellar orbits. The orbits of the IMBH and the test stars precess according to Equation (\ref{eq:tprec}), such that  for each time step, $dt$, they rotate in their orbital plane by an angle $|\delta \phi| = 2 \pi (dt/t_{\rm prec})$. 

\begin{figure}[t!]
	\begin{center}
			\includegraphics[trim=0cm 1.5cm 0cm 0cm, clip=false, height=140 mm, angle=270]{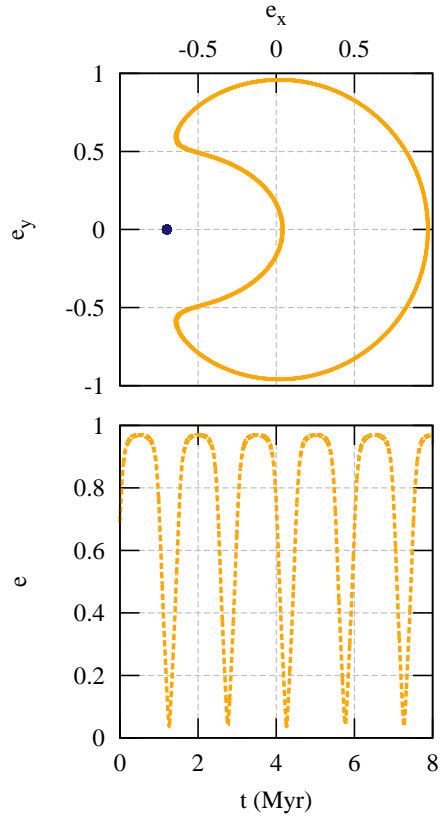}
		\caption{Example of a secular horseshoe orbit of a star of mass $m = 1 \Mo$, bound to an SMBH of mass $\Mbh = 4 \times 10^6 \Mo$. The stellar orbit feels a gravitational torque from an IMBH (mass $\Mimbh = 10^3 \Mo$) and precesses due to general relativity and a background stellar potential. Top: the eccentricity vectors of the IMBH and the prograde-directed single star in the frame co-precessing with the eccentricity vector of the IMBH. Bottom: eccentricity evolution of the stellar orbit over $8 \Myr$ during which it performs several secular horseshoe orbits. 		
		 \label{f:sim_shso}}
	\end{center}
\end{figure}

As the IMBH moves in a smooth stellar potential, it experiences no dynamical friction from it. To account for this we use the dynamical friction formulae given by \citet{Jus11} to artificially decrease the semi-major axis of the IMBH's orbit. \citet{Jus11} use a self-consistent velocity distribution function in place of the standard Maxwellian assumption and an improved formula for the Coulomb logarithm to arrive at dynamical friction timescale for a near-Keplerian potential, which can differ by a factor of three from the standard formula in Equation (\ref{eq:chandra}). 

The radial evolution of the IMBH (point-like object) behaves, for $\alpha \ne 3/2$,  as\footnote{For  $\alpha = 3/2$ see similar equations in \citet{Jus11}.}
\begin{equation} \label{eq:jus10_a}
r = r_0 \left(1 - \frac{t}{t_{\rm DF}} \right)^{\frac{2}{2\alpha - 3}},
\end{equation}
where $t_{\rm DF}$ is the dynamical friction timescale calculated from the starting position of the massive body $r_0$,
\begin{equation}
\label{eq:tdf}
t_{\rm DF} = \frac{P}{2 \pi (3 \!-\! \alpha) \chi \ln \Lambda} \frac{\Mbh^2}{\Mimbh M_<}.
\end{equation} 

$\chi$ is the fraction of stars with velocity smaller than that of the IMBH (circular case),
\begin{equation}
\chi(\alpha) = \frac{1}{\sqrt{2}B(3/2,\alpha\!-\!1/2)} \int_0^1 u^2 \left(1 - \frac{u^2}{2}\right)^{\alpha-3/2} \!du
\end{equation}
and $\Lambda$ is given by
\begin{equation}\label{eq:lambda}
\Lambda = \frac{(3 + \alpha)}{\alpha(1+\alpha)} \frac{\Mbh}{\Mimbh}.
\end{equation}

We implement Equations (\ref{eq:jus10_a}) \-- (\ref{eq:lambda}) by adjusting the semi-major axis of the IMBH orbit at each time step with the following transformation
\begin{equation}
r_t \rightarrow \beta ~ r_{\rm t-1} \quad v_t \rightarrow \frac{v_{\rm t-1}}{\sqrt{\beta}},
\end{equation}
\begin{equation}
\beta =  \left(1 - \frac{t}{t_{\rm DF}(r_{\rm t-1})} \right)^{\frac{2}{2\alpha - 3}}.
\end{equation}
Essentially we are re-scaling the stellar ellipse, without changing its eccentricity. In the next subsections we present three examples of dynamics resulting from SDAF.

\subsection{Secular horseshoe orbits}

Horseshoe orbits describe the motion of satellite bodies in the circular planar restricted three-body problem \citep[for a comprehensive overview see][]{Der99} which librate on paths encompassing the $L3$, $L4$, and $L5$ Lagrangian points. Such orbits map out horseshoe shapes in the frame co-rotating with two massive bodies and have been observed in the Saturnian satellite system \citep{Der81} and in the co-orbital Earth Asteroid 2002 $\rm{AA_{\rm 29} }$\citep{Con02}. 

Secular dynamical anti-friction results in {\it secular} horseshoe orbits.\footnote{Referred to as ``windshield-wiper orbits'' in \citet{Mer11a} in the context of non-axisymmetric star clusters.} Secular horseshoe orbits differ from those in our above description in that it is the eccentricity vector, as opposed to the guiding center of the orbit which maps out a horseshoe in the frame co-rotating with the IMBH \citep[for close analogies in planetary dynamics, see][]{Mit04,Pan04,Lit11}. We illustrate this behavior in Figure \ref{f:sim_shso} where we show the position of the eccentricity vectors of a IMBH and a star in the frame of reference co-precessing with the eccentricity vector of the IMBH. In this example, $\Mbh = 4 \times 10^6 \Mo$, $\Mimbh = 10^3 \Mo$, $m = 1 \Mo$, $\alpha = 1$, $M(<1\pc)=1 \times 10^6 \Mo$, $\aimbh \! = \! a \!= \!0.0703 \pc$, $\eimbh = 0.7$, and $e = 0.69$. Both the IMBH and stellar orbits lie on the $xy$-plane with an initial angle of $\delta \phi = \pi/4 ~\rad$ between their eccentricity vectors. This behaviour is transitory due to non-secular gravitational interactions. The timescale over which the star in this simulation performs a horseshoe in eccentricity space is $\sim1.5 \Myr$ or $\sim1700$ orbital periods. 

\begin{figure}[t!]
  \centering
  \subfigure{
			\includegraphics[trim=0cm 0cm 0cm 1.5cm, clip=true, height=79 mm, angle=270]{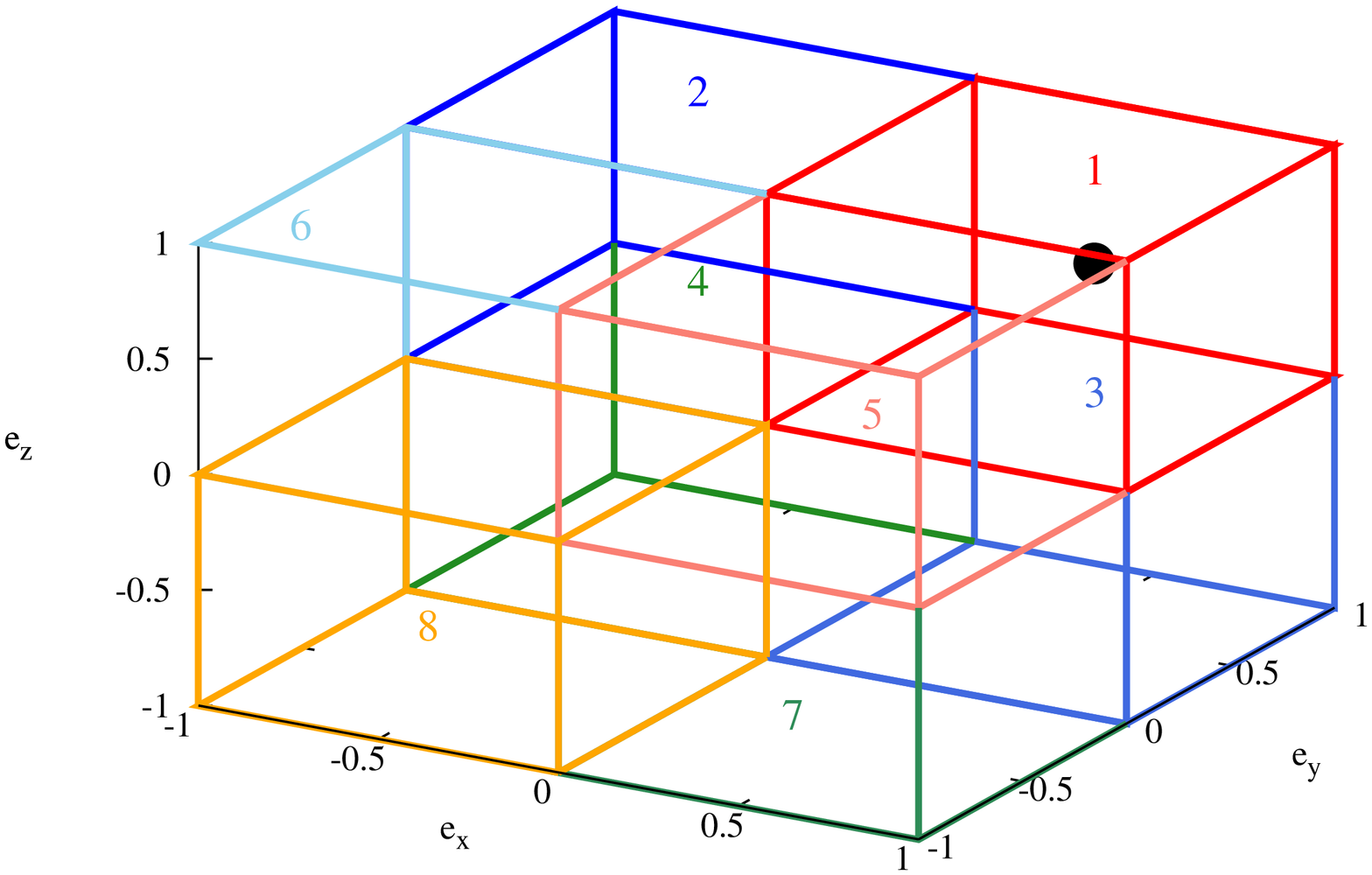}
  }  \subfigure{
			\includegraphics[height=83 mm, angle=270]{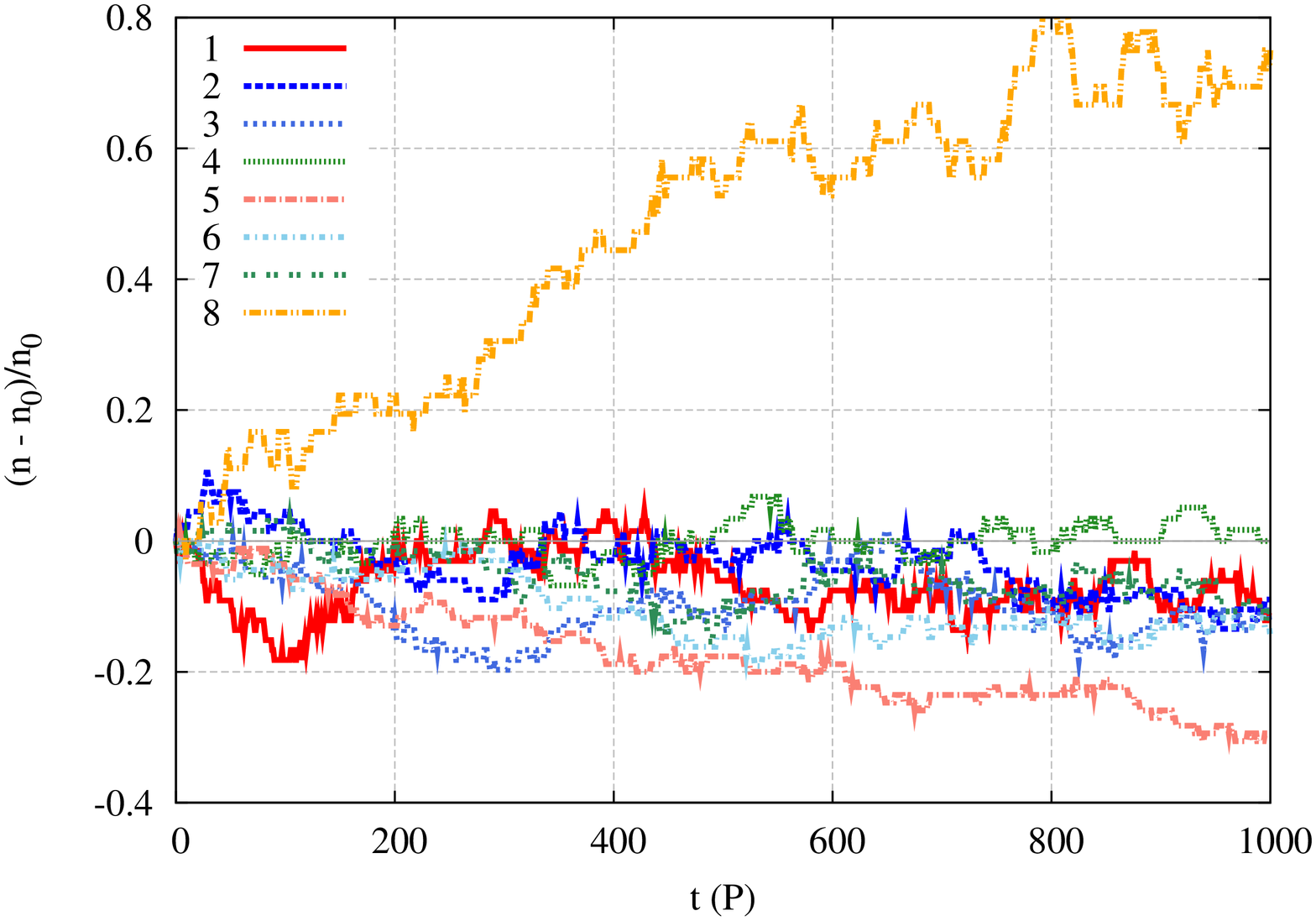}
  }
  \subfigure{
			\includegraphics[height=83 mm, angle=270]{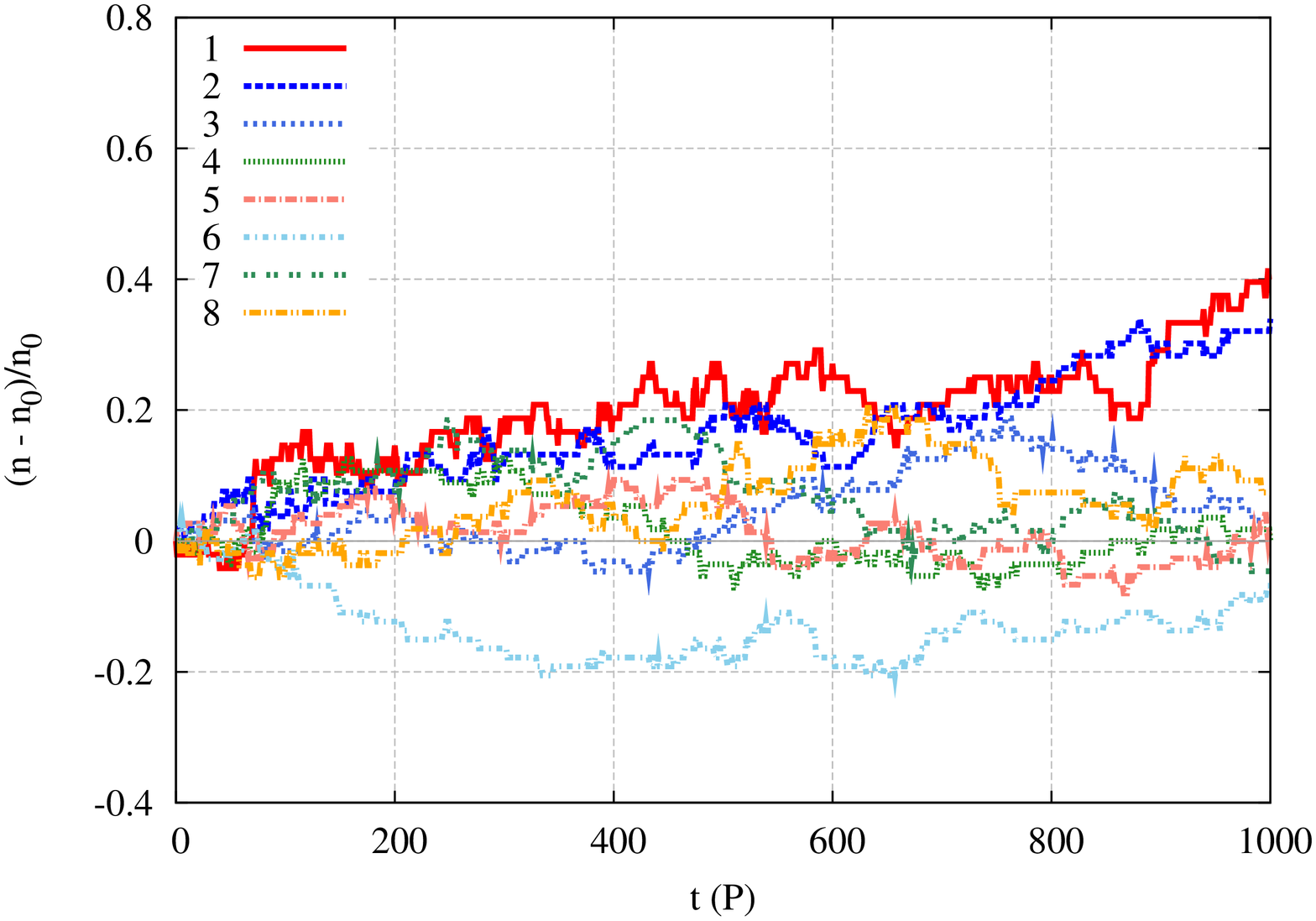}
  }
		\caption{Schematic of octants in eccentricity vector space (top). The unit eccentricity vector of the IMBH lies parallel with (1,1,1) in octant 1 (black dot). Fractional change in number of prograde (middle) and retrograde (bottom) stars in each octant in the frame rotating with the IMBH. Time is given in units of initial orbital period of the IMBH, $P (a = 0.06 \pc)$. The increase in the number of prograde stellar orbits in octant 8, and retrograde stellar orbits in octant 1, is as predicted. 
				 \label{f:quad}}
\end{figure}


\subsection{Separation of potential into pro- and retro- grade orbits}

As illustrated in Figure \ref{f:sim_shso} the eccentricity vectors of prograde orbits which have a small inclination angle with respect to an IMBH will tend to librate about a point $\pi ~\rad$ opposite to that of the IMBH \--- the secular analog of the Lagrangian $L3$ point. In contrast, the eccentricity vectors of retrograde orbits (those which do not flip to prograde orbits) will, in a time-averaged sense, cluster around that of the IMBH before they are pushed away, as they precess more slowly due to the increase in their orbital eccentricity. We find that the presence of a massive body, such as an IMBH, surrounded by less massive objects in a near-Keplerian potential tends to segregate the lighter objects into a region dominated by prograde orbits and one dominated by retrograde. 

This mechanism has the strongest effect for, but is not restricted to, co-planar orbits, as scalar angular momentum changes result in eccentricity changes whereas vector angular momentum relaxation re-orients the stellar orbital plane. 

In Figure \ref{f:quad} we plot an example of this segregation. The top figure shows a schematic of eight octants in eccentricity vector space. We rotate the cluster such that the eccentricity vector of the IMBH lies parallel with (1,1,1) and follow the change in the number of massless stellar orbits with eccentricity vectors in each octant. The middle plot shows the results for prograde orbits and the bottom plot for retrograde. There is a clear increase in the number of prograde stellar orbits in octant 8, opposite that in which the IMBH lies, as expected. Furthermore, there is an enhancement of retrograde stellar orbits in octant 1 as the theory predicts.

\subsection{Increase of orbital eccentricity of IMBH}

Here we examine two of the main testable predictions of SDAF: the eccentricity evolution of an IMBH with (1) artificially reversed apsidal precession and (2) with a decreased migration rate from Chandrasekhar's dynamical friction. We present the results of short-term simulations performed to test these predictions, using ten realizations for each simulation. In the following, $\Mbh = 4 \times 10^6 \Mo$, $\Mimbh = 10^4 \Mo$, $\aimbh \! = \!0.06 \pc$, $\eimbh\!  = \! 0.4$, $\alpha \! = \! 1.5$, and $M(\! <\! 1\pc)=1 \times 10^6 \Mo$. We integrate the orbits of $1000$ test stars in each simulation, distributed from $0.006 \pc \! < \! a \! < \! 0.12 \pc$, with $m = 20 \Mo$. 

\begin{figure}[t!]
  \centering
			\includegraphics[height=90 mm, angle=270]{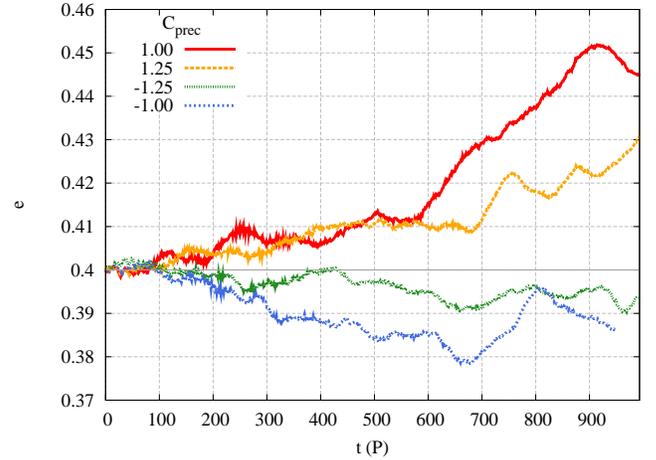}
 		\caption{Mean eccentricity evolution (each curve is the average of ten simulation runs) of an IMBH experiencing four different rates of apsidal precession. The first is the reference simulation with the correct precession rate corresponding to the smooth background potential (red solid line). In the second simulation, the IMBH has an increased precession rate by a factor of $1.25$ (orange dotted line). In the third simulation, the IMBH has an increased precession rate by the same factor but the IMBH is made to precess in a {\it reversed} direction (green dotted line). In the fourth simulation, the IMBH has a precession period corresponding to the reference simulation but again in a reversed direction (blue dotted line). Time is given in units of the initial orbital period $P (a = 0.06 \pc)$.	
				 \label{f:prec}}
\end{figure}

In Figure \ref{f:prec}, we show the evolution of orbital eccentricity of the IMBH as a function of time in units of the initial orbital period, $P (a \!= \! 0.06 \pc) \! \simeq \! 692 \yr$. Each line shows the mean eccentricity of ten simulations. $C_{\rm prec}$ is a pre-factor placed in front of the apsidal precession angle, $\delta \phi$, at each time step. The reference simulation with the correct precession rate $C_{\rm prec} \!  =\!  1.00$ shows an increase in eccentricity as expected. Increasing the precession rate by a factor of $1.25$ does not however increase the rate of eccentricity growth, as the IMBH orbit has less time to exert torques on the surrounding stellar orbits. In the next two simulations we reverse the direction of apsidal precession of the IMBH, keeping the precession period roughly constant in relation to the first two simulations. The overall eccentricity growth of the IMBH is negative, as predicted. 

\begin{figure}[t!]
  \centering
			\includegraphics[height=90 mm, angle=270]{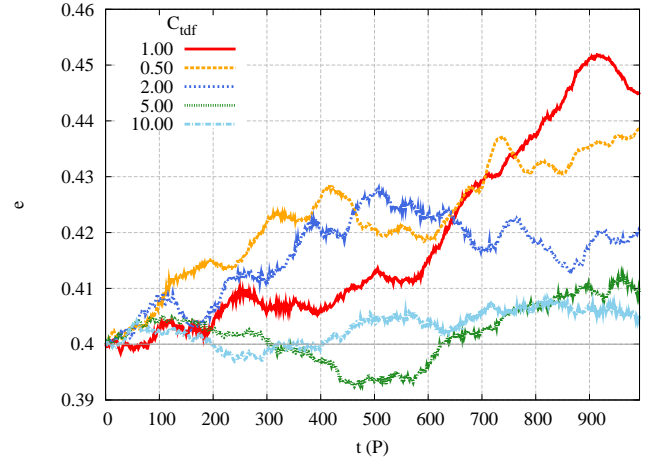}
 		\caption{Mean eccentricity evolution (each curve is the average of ten simulation runs) of an IMBH with five different migration rates due to dynamical friction. SDAF predicts lower rates of eccentricity growth with increasing dynamical friction timescales. The parameter $C_{\rm tdf}$ is a multiplicative pre-factor in front of the dynamical friction timescale given in Equation (\ref{eq:tdf}). 				 \label{f:mig}}
\end{figure}

Next, we present the results of simulations wherein we change the migration rate of the IMBH (Figure \ref{f:mig}). SDAF predicts that lower migration rates lead to slower eccentricity growth as the IMBH interacts with retrograde orbits at a lower rate. In this figure, $C_{\rm tdf}$ is a multiplicative pre-factor in front of the dynamical friction timescale given in Equation (\ref{eq:tdf}). The semi-major axis of the IMBH orbit shrinks from $0.06 \pc$ to $\sim0.024 \pc$ over the simulation for $C_{\rm tdf} = 1$. For $C_{\rm tdf} \!=\! 2, 5, 10$ the eccentricity growth of the IMBH is suppressed (though still positive), exactly as SDAF predicts. In the simulation in which we half the friction timescale ($C_{\rm tdf} = 0.5$), and hence increase the migration rate by a factor of two, the eccentricity growth rate does not increase. We attribute this to the shortening over the timescale over which the IMBH exerts persistent torques on the surrounding stellar orbits.

Finally, we run simulations wherein we soften the IMBH's gravity to such an extent that it switches off entirely for a star coming within $0.01 \pc$ of the IMBH itself. We do so to confirm that secular dynamics are dominating the eccentricity evolution of the IMBH orbit; these conditions inhibit the three-body scatterings and ejections which are the basis for the theories given in \citet{Iwa11} and \citet{Ses11}. We find it makes no significant difference for the angular momentum evolution of the IMBH even though it significantly reduces the number of stellar ejections (by a factor of $\sim 5$). 

\pagebreak

The problem of dynamical friction on a bar rotating through a spherical stellar system has been studied by \citet{Tre84}. The authors find that the torque is directed opposite to the direction of the bar's rotation; see, e.g., their Equation (67). This seems in contradiction with our finding that the precessing IMBH orbit experiences a torque directed in the same direction as its precession. We note however, that Equation (67) of \citet{Tre84} is derived under the assumption that the bar's mass is small enough such that its effect on the orbits of the stars is weak and perturbation theory can be used. We, on the other hand, explore the regime where the relevant secular interactions between the IMBH and the stars are strong. A secular horse-shoe orbit is an example of such an interaction, an angular momentum flip is another one.

\subsection{Eccentricity evolution in a rotating stellar cluster} 

For a non-rotating stellar cusp, the eccentricity increase in the IMBH orbit due to SDAF is relatively small (as seen in Figures \ref{f:prec} and \ref{f:mig}). However, net rotation is observed in most well-resolved galactic nuclei \citep[e.g.,][]{Set10b}, including our own \citep{Gen96}.
In this case, there is a higher rate of strong secular encounters between either retrograde or prograde stellar orbits and that of the IMBH and, as a result, the rate of eccentricity evolution of the IMBH orbit due to SDAF is greatly enhanced. Specifically, SDAF predicts a strong increase in orbital eccentricity of the IMBH in counter-rotating clusters due to the increased number of retrograde stellar orbits, while it predicts a decrease for co-rotating clusters due to the increase in prograde orbits.

\begin{figure}[t!]
  \centering
			\includegraphics[height=90 mm, angle=270]{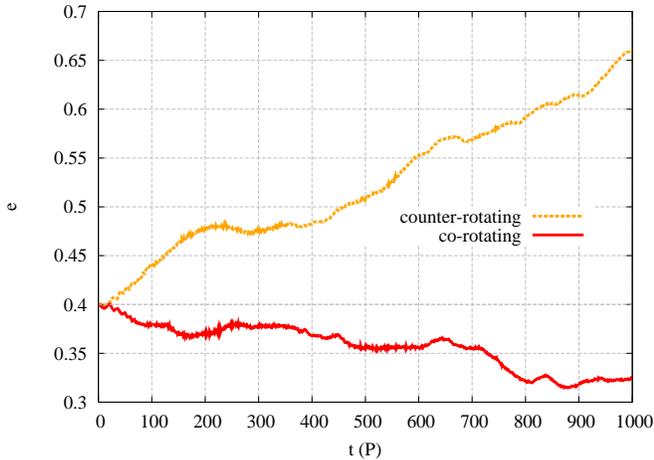}
 		\caption{Mean eccentricity evolution (each curve is the average of ten simulation runs) of an IMBH orbiting in counter- and co-rotating stellar cusps. The evolution is faster than in non-rotating stellar cusps. 				 \label{f:rot}}
\end{figure}

 In Figure 7, we show the results of simulations with co- and counter-rotating stellar cusps to demonstrate the significant change in eccentricity of the IMBH above that achieved in non-rotating cusps. Using the same parameters as in our previous simulations, we see a mean increase (decrease) in eccentricity of $e = 0.4$ to $e \approx 0.66$ ($0.32$) in the counter- (co-) rotating stellar cusp scenario. Demonstrating a more rapid growth in eccentricity requires additional ``grainy'' mass in surrounding cusp stars (i.e., $N$-body/field particles as opposed to a smooth potential). We refer the reader to Figure 2 of \citet{Ses11} who simulate a similar potential to that in our model but use significantly more $N$-body particles. They find a growth (decrease) in eccentricity of e = $0.5$ to $e \approx 1$ ($0.06$) over a similar timescale. However, in contrast to \citeauthor{Ses11}'s interpretation, we attribute this growth (decrease) to SDAF.  

\section{Discussion}

We have presented a new secular gravitational-dynamical process in galactic nuclei which acts to decrease the orbital angular momentum of a massive body. There are two essential components to the decrease. The first is the net flow of retrograde orbits past that of the massive body which extracts orbital angular momentum. The second is Chandrasekhar's dynamical friction force acting on the massive body which decreases its semi-major axis and provides a mechanism for net negative angular momentum change. Although we have told this story in the context of an IMBH inspiral into the center of a galaxy, the interactions described here will be valid for any massive body in a near-Keplerian potential. 

SDAF can account for the rise in eccentricity only in the initial stages of the inspiral of the IMBH, not during the stalling phase. The eccentricity does rise strongly in the stalling phase, as was identified in  \citet{Bau06} and \citet{Mat07}. However, the origin of this increase is not SDAF (which does not occur for non-migrating orbits), but is instead due to the conventional dynamical friction being reduced at the periastron (inside the cavity) as compared to that at the apoastron (outside the cavity or closer to its edge). This effect was clearly explained in \citet{Mat07} and for IMBHs it plays decisive role. SDAF does however play an important role if the cusp is co- or counter-rotating relative to the IMBH's orbit.

Observational predictions of SDAF include higher mass objects with increasingly larger eccentricities (in non-rotating and counter-rotating stellar cusps), and density anisotropy in the stellar distribution near the plane of the massive body.

\acknowledgments{A.-M.M gratefully acknowledges the hospitality of Monash University and the Aspen Center for Physics. We thank Fabio Antonini, Dan Fabrycky, Clovis Hopman, Margaret Pan, Alberto Sesana and Scott Tremaine for valuable discussions and/or comments on the text. We thank Holger Baumgardt, the referee, for several good comments and suggestions that helped improve
the paper. A.-M.M is supported by a TopTalent fellowship from the Netherlands Organisation for Scientific Research (NWO) and Y.L. by a VIDI grant from NWO.}

\end{document}